\documentclass[fleqn,twoside]{article}
\usepackage{espcrc2}

\usepackage{graphicx}
\usepackage[figuresright]{rotating}

\title{Imaging Protoplanetary Disks with a Square Kilometer Array}

\author{D.J. Wilner\address[CfA]{
        Harvard-Smithsonian Center for Astrophysics,
        60 Garden St., Cambridge, MA, 02138 USA}}
       
\begin{document}

\begin{abstract}
The recent detections of extrasolar giant planets has revealed a
surprising diversity of planetary system architectures, with many very
unlike our Solar System. Understanding the origin of this diversity
requires multi-wavelength studies of the structure and evolution of
the protoplanetary disks that surround young stars.  Radio astronomy
and the Square Kilometer Array will play a unique role in these
studies by imaging thermal dust emission in a representative sample of
protoplanetary disks at unprecedented sub-AU scales in the innermost
regions, including the ``habitable zone'' that lies within a few AU of
the central stars.  Radio observations will probe the evolution of
dust grains up to centimeter-sized ``pebbles'', the critical first
step in assembling giant planet cores and terrestrial planets, through
the wavelength dependence of dust emissivity, which provides a
diagnostic of particle size.  High resolution images of dust emission
will show directly mass concentrations and features in disk surface
density related to planet building, in particular the radial gaps
opened by tidal interactions between planets and disks, and spiral
waves driven by embedded protoplanets.  Moreover, because orbital
timescales are short in the inner disk, synoptic studies over months
and years will show proper motions and allow for the tracking of
secular changes in disk structure.  SKA imaging of protoplanetary
disks will reach into the realm of rocky planets for the first time,
and they will help clarify the effects of the formation of giant
planets on their terrestrial counterparts.
\vspace{1pc}
\end{abstract}

\maketitle

\section{Introduction}
The detection of extrasolar planets has fired the imagination of
modern society to the possibility of finding other Earth-like planets.
The search for terrestrial exoplanets that could harbor life is one of
the most publicly enthralling and scientifically and philosophically
important ventures of the 21st century. The Square Kilometer Array
(SKA) will make a fundamental and crucial contribution to this effort
as the only observational facility currently planned that will be
capable of imaging the birth sites of terrestrial planets in the dusty
disks surrounding newly formed stars-- the``cradle of life''.

The dusty disks that naturally arise from the star formation process
are the sites where planets are made\cite{beckwith96}.  Gravitational,
hydrodynamic, magnetic, and chemical processes are all active in
determining the evolution of disks and their planet forming activity.
The inner disk, within a few AU of the central star, is the
``habitable zone'' for stars like the Sun, within which terrestrial
planets or the moons of gas giants are most likely to have
environments favourable for the development of life. The SKA will have
the unique capability to image thermal dust emission at unprecedented
sub-AU scales within this zone.  The surprising discovery of giant
exoplanets located well within the habitable zone raises many
questions: What accounts for the diversity in planetary systems?  Are
terrestrial planets common in the habitable zone?  Do giant planets
form in the inner disk or do they migrate there, and what are the
implications for terrestrial planets?  Is our Solar System an
exceptional case?  The answers to these questions will come from a
better understanding of the structure and evolution of protoplanetary
disks.

\section{Protoplanetary Disks} 
\subsection{Observational Challenges}
By the time stars of $\sim1~$M$_{\odot}$ reach ages of $\sim1$~Myr, 
the majority of their natal circumstellar envelope material has been 
dispersed and they are revealed at optical wavelengths as T-Tauri stars.  
At this stage, the accretion disks surrounding the stars where planets 
form become readily accessible to observation at many wavelengths.
These observations face several significant challenges:
\begin{itemize}

\item The bulk of the disk mass is comprised of molecular hydrogen
that does not emit at the low temperatures characteristic of the
material.  The disk can be studied only through its minor
constituents, in particular (1) dust grains, through thermal emission
and scattered starlight, and (2) trace molecules with fractional
abundances of $10^{-4}$ or less, many of which are frozen out onto
dust grains except in the heated upper layers of disk atmospheres
where they participate in a complex chemistry.

\item
The angular scales of even nearby disks are small. For the large
sample of $\sim1$~Myr old stars associated with the nearest dark
clouds like Taurus, Ophiucus and Chamaeleon at $\sim150$~pc, the 10
and 2 AU diameters of the orbits of Jupiter and Earth subtend only 65
and 13 milliarcseconds respectively. Thus very high angular resolution
is required to discern disk structures at the size scales of interest.

\item 
The inner disk, especially within a few AU of the central star, is
difficult to probe at most wavelengths even when sufficient angular
resolution may be obtained.  In the optical, the high contrast of
scattered light from the disk with the stellar photosphere is
problematic, and very careful point-spread-function subtraction or
coronography is essential. Even with these techniques, the habitable
zone remains inaccessible. At millimeter wavelengths, dust emission
remains optically thick at the typical column densities of
$100$~g~cm$^{-2}$ or more, and observations cannot penetrate into the
disk to reveal structure.

\end{itemize}

\subsection{The Role of the SKA}
Though the investigation of protoplanetary disk environments has not
been a traditional topic for radio astronomy, the SKA will uniquely
meet the challenges of imaging the inner regions of protoplanetary
disks.  Significant collecting area on $\sim$1000 km scales operating
at wavelengths $\sim22$~GHz ($\sim1.3$~cm) will provide sufficient
sensitivity for imaging thermal dust emission at $\sim$milliarcsecond
resolution.  The short centimeter wavelengths accessible to the SKA
are particularly advantageous at this high angular resolution since
the dust emission remains optically thin and samples the high surface
density regions of the inner disks that remain opaque at millimeter
wavelengths.  Also, emission from circumstellar dust dominates
emission from the stellar photosphere, generally by many orders of
magnitude at these wavelengths, and there is no contrast problem
between disk and star.

SKA images of inner disks will have angular resolution an order of
magnitude higher than for any other planned observational facility,
including the Thirty Metre Telescope at optical wavelengths, the James
Webb Space Telescope in the infrared, and the Atacama Large Millimeter
Array.  This order of magnitude in resolution is an especially
significant one, as it enables imaging of structure in the
habitable zone of nearby protoplanetary disks for the first time.

\subsection{Global Properties}
\label{sec:background}
\begin{figure*}
\begin{center}
\includegraphics[scale=1.25]{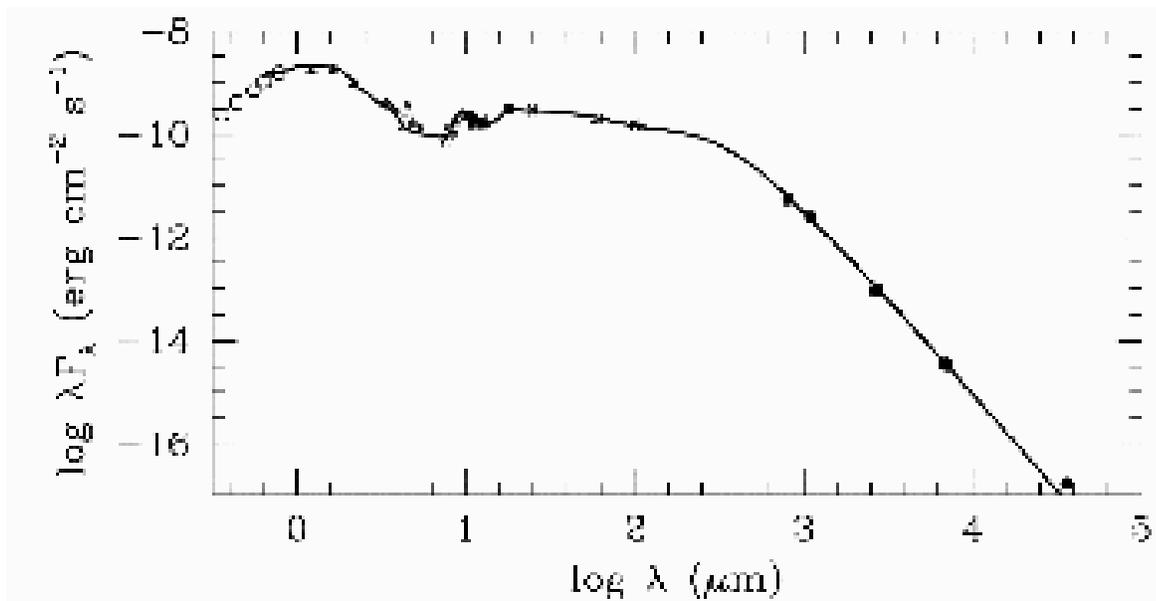}
\end{center}
\caption{The spectral energy distribution of the nearby young star
TW Hya (from \cite{calvet02}). At optical wavelengths, the stellar
photosphere dominates. At mid-infrared and longer wavelengths, the
broad continuous emission comes from dust particles in a circumstellar
disk heated primarily by starlight impinging on the disk surface.
The panchromatic nature of the emission is due to the broad range of
temperatures in the disk, which has radial and vertical gradients.} 
\label{fig:twhya_sed}
\end{figure*}

Since most of the disk material surrounding low mass stars beyond a 
few stellar radii is at low temperatures, well below 1000~K, dust 
provides the dominant source of continuous opacity. Though a large fraction 
of the disk energy is radiated in the far-infrared, a spectral range 
difficult to access from the ground and lacking large apertures in space,
a lot has been learnt about the global properties of protoplanetary 
disks from analyses of their overall spectral energy distributions,
which are sensitive to the size, shape, and mass of the disk. 
Figure~\ref{fig:twhya_sed} shows the spectral energy distribution 
of TW Hya, a nearby T-Tauri star of spectral type K7, from optical to 
radio wavelengths, together with a model of disk emission.

The shape of the spectral energy distribution of TW Hya is fit very well
by the disk model. The spectral energy distributions of most 
T-Tauri stars are generally explained very well by disk models with
a parameterized dust absorption coefficient of power law 
form $\kappa_\nu = \kappa_d (\nu/10^{12}~{\rm Hz})^{\beta}$, with  
$\beta$ $\sim1$ and normalization $\kappa_d=0.1$ cm$^2$~g$^{-1}$ 
(see \cite{beckwith90}).
This expression hides many uncertainties associated with grain sizes, 
composition, and also the dust-to-gas ratio, e.g. the emissivity 
must be affected by the evaporation of different grain constituents, 
starting with water ices at $T\sim200$~K. Nonetheless, this expression
is commonly adopted and provides a useful first-order description.  

The dust emission from a typical protoplanetary disk starts 
to become optically thin at wavelengths longward of a few hundred microns,
which becomes readily accessible to observation through atmospheric windows 
at millimeter and radio wavelengths. In this regime, where all but the
inner disk regions are optically thin, the observed emission is directly 
proportional to disk mass, weighted by temperature.
Early analyses of spectral energy distributions of T-Tauri stars like
TW Hya provided strong support for disk geometries, suggesting
minimum outer radii of 10's to 100's of AU \cite{adams87}
and masses of gas + dust from 0.001 to 0.1~M$_{\odot}$,
sufficient to form planetary systems like our own \cite{beckwith90}. 
The reasons for the wide range in protoplanetary disk properties is not 
known, but it is likely that initial conditions, such as the angular 
momentum content of the parent molecular core, as well as environmental 
effects, are important. 

Statistical studies of large samples suggest 
that disks dissipate on timescales of order 10 Myr 
(e.g. \cite{strom89,haisch01}),
compatible with the standard view 
of giant planet formation by dust coagulation, planetesimal formation, 
and core accretion of nebular gas, as indicated by the fossil record
of the planets in our Solar System \cite{wuchterl00}.
This rough outline of the planet formation process is certainly 
incomplete.
For example, the core accretion model has trouble accounting 
for the formation of ice giant planets at the distances of Uranus and 
Neptune on timescales shorter than the disk dissipation time.  
An alternative model postulates that gas giant protoplanets form very 
rapidly (in just $\sim10^3$ years) by disk instability, where the gas in 
a marginally gravitationally unstable disk forms clumps that then 
contract to planetary densities \cite{boss97}. In addition, 
the discovery that extrasolar giant planets at $<$0.05 AU from their 
parent stars are common, together with their distributions in radius 
and eccentricity, indicates that giant planets likely migrate inward 
from formation zones farther out in the disk (e.g. \cite{lin96}).
The dynamical evolution of these migrating giant planets depends sensitively 
on the physical properties of the disk, in particular the 
surface density structure and the gas dissipation time \cite{goldreich03}.
These dynamical mechanisms have important consequences for the formation
and survival of terrestrial planets in the inner disk, 
which must arise from collisional accumulation of rocky planetesimals.

Spatially unresolved panchromatic observations of the spectral energy
distributions of protoplanetary disks will continue to lead to new
insights, but direct imaging is ultimately essential to form a
complete picture, to verify theoretical constructs, and to break model
degeneracies.

\section{SKA Sensitivity and Imaging Feasibility}
Dust emissivity drops steeply toward long wavelengths, and thermal 
emission from dust at the high angular resolution appropriate to 
the inner regions of protoplanetary disks is very weak by the
standards of today. Remarkably, the SKA will be sufficiently 
sensitive to image dust emission with brightness of $\sim100$~K 
at milliarcsecond resolution in very reasonable integration times. 

For a geometrically thin disk, the flux $dS$ from a dusty disk 
element filling solid angle $d\Omega$ is 
\begin{equation}
dS = B_{\nu}(T) (1-e^{-\tau_{\nu}})\cos{i} \, d\Omega
\end{equation}
where $B_{\nu}(T)$ is the Planck function, $i$ is the inclination,
and the optical depth $\tau_{\nu}$ is given by
\begin{equation}
\tau_{\nu} = \Sigma\kappa_{\nu}/\cos{i}
\end{equation}
where $\Sigma$ is the surface density and
$\kappa_{\nu}$ is the dust mass opacity.
To make an estimate for the SKA, we assume an observing frequency 
of 22~GHz, a conservative fiducial value that represents a compromise 
between the rapid 
increase in flux with frequency, the increase with opacity with frequency,
and the likely decrease in SKA sensitivity with frequency. 
The optimal choice of observing frequency will depend on details of the 
system performance and the distribution of collecting area with 
baseline length, and it is likely to be in the range of 22 to 34 GHz. 
With these assumptions, the disk optical depth is 
\begin{equation}
\tau_{\nu} = 0.31 \left(\frac{\nu}{22~{\rm GHz}}\right)
            \left(\frac{\Sigma}{100~{\rm g/cm^2}}\right)
            \left(\frac{\cos i}{\cos 45}\right)^{-1}
\end{equation}
and, for low optical depth in the Rayleigh-Jeans regime
\begin{eqnarray}
dS & = & 0.11~{\mu{\rm Jy}} \left( \frac{T}{300~{\rm K}} \right)
           \left( \frac{\Sigma}{100~{\rm g/cm^2}} \right) \\ 
~ & ~ & \times \left( \frac{\nu}{22~{\rm GHz}} \right) ^3
           \left( \frac{\theta}{2~{\rm mas}} \right) ^2 \nonumber
\end{eqnarray}
where $\theta$ is the synthesized beam Gaussian FWHM size.
At 22 GHz, an angular resolution of 2 milliarcsconds requires baseline 
lengths of $\sim1500$~km. 
If the SKA reaches an rms sensitivity of $\sim0.02$~$\mu$Jy in 8~hours,
as expected, then dust emission from the inner regions of disks where 
the brightness temperature exceeds 100~K and the optical depths approach unity 
will be detectable in hours, and it may be imaged with high signal-to-noise 
in a few tens of hours. Since brightness temperature sensitivity is 
proportional to synthesized beam area, a survey of many sources could be 
made at somewhat lower angular resolution to identify those of most interest 
for the highest resolution imaging in a small fraction of this time 
per source. Note that the typical 22~GHz flux integrated over an entire 
protoplanetary disk spanning a few arcseconds on the sky is of order 
$\sim100$~$\mu$Jy.

For sources at $\sim150$~pc, a flux level of $\sim0.1$~$\mu$Jy at 
22~GHz corresponds to less than an Earth mass of an interstellar mixture 
of gas and dust filling a 2~milliarcsecond beam, using the mass opacity 
adopted here.  Thus the SKA will be able to discern mass concentrations 
that could be the seeds of giant planets that will form via the
gravitational instability mechanism, as well as the spiral density waves 
of modest contrast expected to be driven by otherwise 
invisible embedded giant protoplanets and lower mass protoplanets.

Observations at 22 to 34 GHz with the SKA will be also sensitive to 
small amounts of ionized gas in the star-disk system, if present. 
A plasma could contaminate the signal from dust emission, though
any contamination can be estimated accurately via extrapolation of 
longer wavelength data and/or recognized from its spatial distribution, 
which will be different from the dust that fills the disk.  
For example, some T-Tauri stars are known to have chromospheric activity, 
which will be clearly indicated by a high brightness signal localized to 
transient magnetic features comparable in extent to the star. It's
conceivable that additional compact emission features like this 
may prove helpful for calibration purposes on the longest baselines.

\section{Grain Growth Studies}
\begin{figure}
\begin{center}
\includegraphics[scale=0.45]{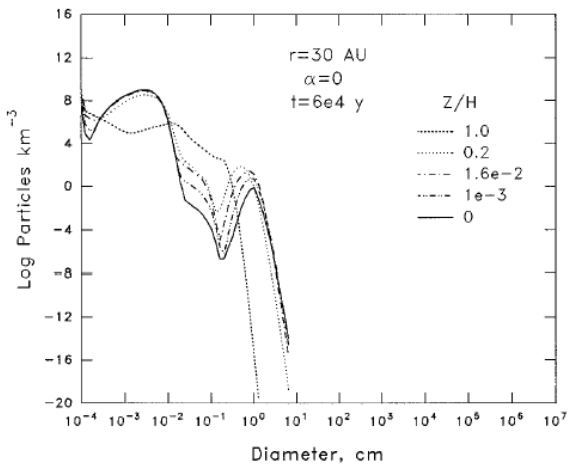}
\end{center}
\vskip -0.75cm
\caption{A numerical simulation of particle size evolution within
a protoplanetary disk (from \cite{weiden97}). The small grains
in the disk stick and settle to the mid-plane, which creates a layer 
of centimeter size particles. The plot shows the particle size
distribution at different elevations, with the mid-plane that contains
most of the mass indicated by a solid line. 
}
\label{fig:weiden97}
\end{figure}

One mysterious aspect of the planet building process is how sub-micron
size interstellar dust grains overcome energetic obstacles that
prevent sticking, grow to sizes large enough to decouple from the
disk gas, interact gravitationally, and build up to planetesimal
sizes.  This is a critical first step in the development of both giant
planet cores and rocky planets.  Most models postulate a phase of
collisional agglomeration mediated by mechanisms like Brownian motion
and turbulence. Numerical simulations of dust evolution indicate a
bimodal distribution of particle sizes develops as centimeter-size
aggregates grow and settle to the mid-plane
(e.g. \cite{weiden97}). The resulting dense layer becomes the
reservoir for the formation of larger bodies.
Figure~\ref{fig:weiden97} shows the particle size distributions as a
function of height from one such calculation; note the large
population of centimeter-sized grains in the mid-plane.

The wavelength dependence of the dust emissivity provides a diagnostic
of particle size, and dust in many protoplanetary disks shows evidence
for size evolution from a primordial interstellar distribution
\cite{beckwith91,beckwith00}. There is ample evidence that the spectral
slope of dust emission from T-Tauri stars in the millimeter regime is
significantly lower than in diffuse molecular clouds, where the
power-law exponent of the dust emissivity $\beta=2$.  For example, the
spectral slope of the TW Hya disk shows a shallow dependence of
emissivity on frequency between 1.3~mm (230 GHz) and 7~mm (43~GHz),
with $\beta\sim0.7$ (and resolved images confirm that optical depth
effects do not affect this determination).  For compact spherical
particles of size $<< \lambda$, $\beta=2$, while for particles of size
$>> \lambda$, $\beta=0$.  In realistic astrophysical mixtures, the
value of $\beta$ also depends on grain composition, structure, and
topology, though the particle size generally dominates, especially for
common silicates \cite{pollack94}.  A substantial mass fraction of the
dust grains in protoplanetary disks has apparently evolved and started
to agglomerate.  The low values of $\beta$ derived from observations
of TW~Hya and a few other stars are robustly interpreted as evidence
for particle growth to sizes of order a millimeter or more
\cite{calvet02,natta04}.

In general, observations are sensitive to particles with sizes of
less than a few wavelengths, since the absorption cross section 
cannot depart significantly from the geometric cross section.
The SKA, by observing dust emission at short centimeter wavelengths, 
can probe the presence of particles larger than a millimeter,
a critical regime.  Historically, inadequate sensitivity and spatial 
resolution have thwarted the goal of detecting these ``pebbles'',
since the opacity per unit mass decreases for larger particles and 
the resulting emission is weak. The SKA will allow for the first time 
multi-wavelength imaging that will localize regions within the disks 
with different 
spectral signatures and therefore different grain properties. The observed 
spatial variations in resolved images, together with dust models, 
will shed light on the roles of settling and turbulent entrainment 
in the grain growth process. Resolved observations at centimeter
wavelengths will address the argument that grain growth to sizes
of at least 1~cm are necessary before particles start settling 
to the mid-plane because smaller grains are stirred by turbulence
\cite{weiden00}.

\begin{figure*}
\begin{center}
\includegraphics[angle=270,scale=0.75]{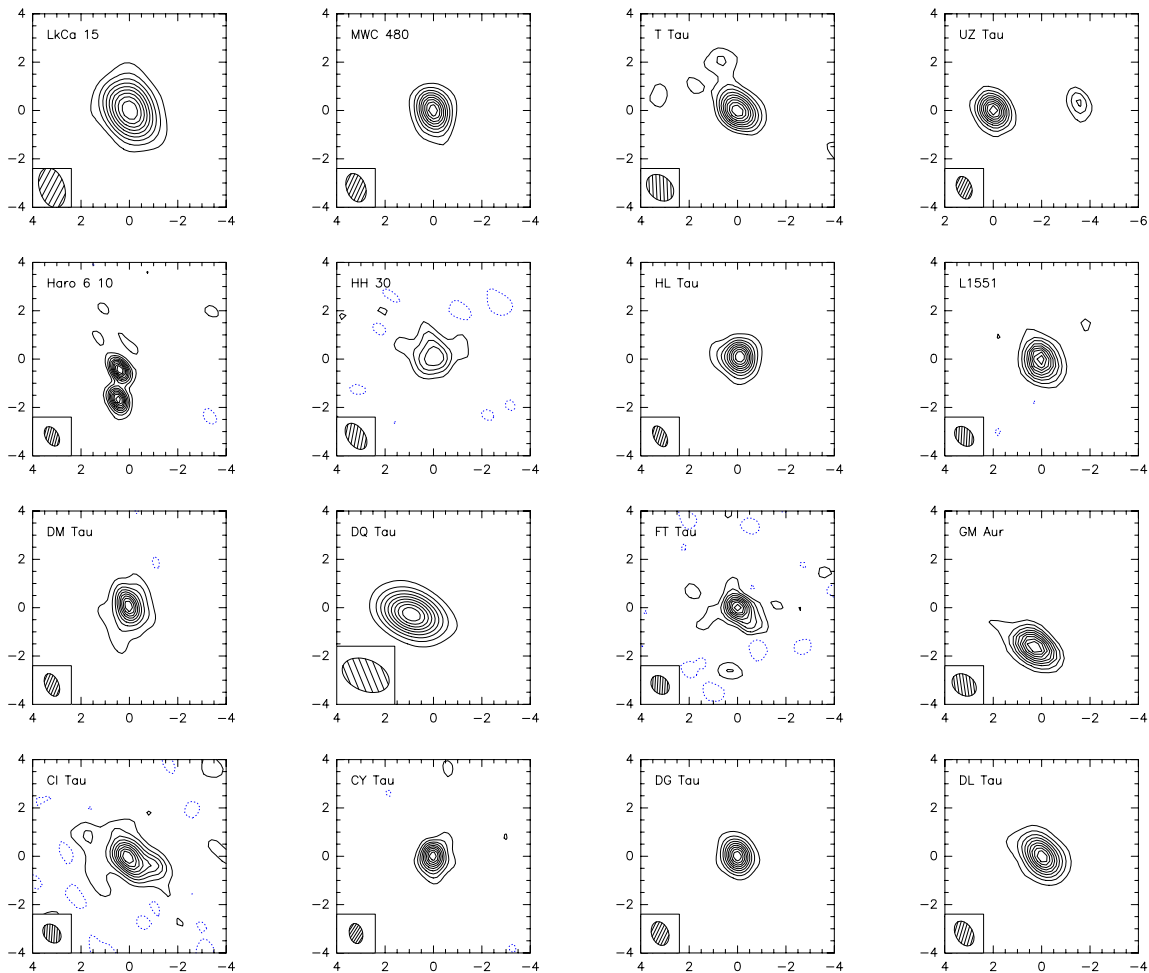}
\end{center}
\caption{Survey of Taurus pre-main-sequence stars (single and multiple) 
from the IRAM Plateau de Bure Interferometer at 230~GHz with typical 
resolution $0.\!\!^{\prime\prime}5$, or 70 AU (from \cite{dutrey01}).
Contours are 10\% of the peak. The optically thin dust brightness from
the circumstellar disks falls steeply with radius in the outer disk,
and it rapidly falls below the sensitivity threshold. Nonetheless, most
of the disks are resolved, and these images provide useful limits on 
the disk size, orientation, and overall surface density structure.
The inner disk regions of planet formation are not probed, primarily 
because of beam dilution. The SKA will push imaging of thermal dust
emission in the high column density inner disks to sub-AU scales, at
lower frequencies where dust opacity is low enough to penetrate through
the full disk column.
}
\label{fig:dutrey_pdbi}
\end{figure*}

\begin{figure*}
\begin{center}
\includegraphics[scale=0.65]{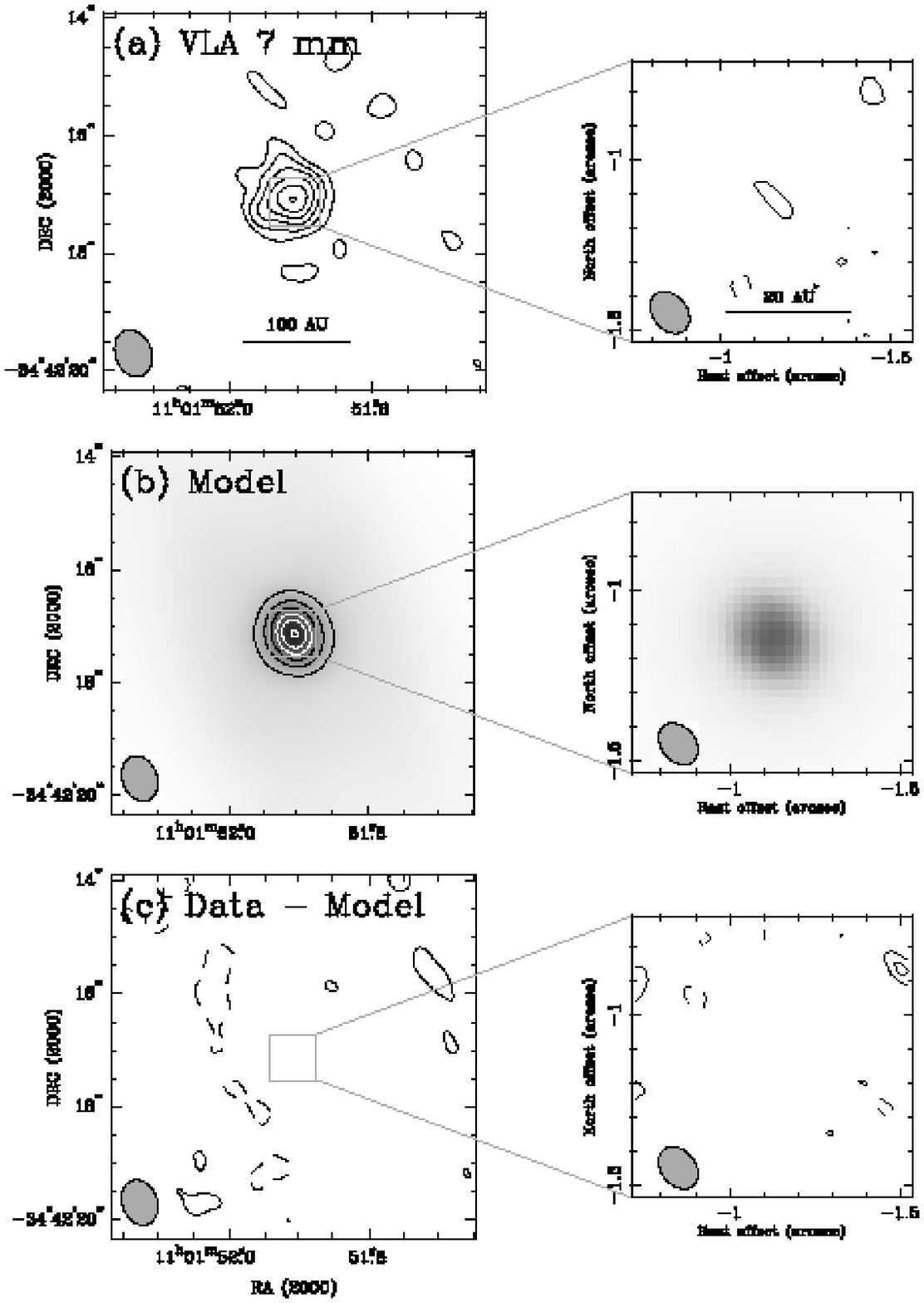}
\end{center}
\caption{
VLA 43~GHz images of TW Hya compared to the accretion disk model that
fits the spectral energy distribution shown in Figure~\ref{fig:twhya_sed}
(from \cite{calvet02}). 
(a) VLA 43 GHz images of TW Hya at two resolutions. 
(b) Simulated VLA images of the disk model brightness distribution with
the same contours and beam sizes as in (a). The gray-scale shows extended 
emission, most of which remains undetectable with the available sensitivity. 
(c) Difference images obtained by subtracting visibilities derived from 
the model from the data. 
}
\label{fig:twhya_vla}
\end{figure*}

\section{Disk Structure: State-of-the-Art}
While current observational capabilities are a far cry from the
potential of the SKA, the results and their interpretation offer a
glimpse at the possibilities for future observations of protoplanetary
disks with higher angular resolution and sensitivity.  The currently
operating millimeter arrays routinely detect large numbers of nearby
protoplanetary disks, and the ongoing upgrades to shorter wavelengths
and longer baselines are just starting to provide spatially resolved
information.  With considerable effort, several interferometric
imaging {\em surveys} of tens of T Tauri stars with
$0.\!\!^{\prime\prime}5$ resolution have been completed, and a few
observations have pushed the angular resolution frontier to
$\sim0.\!\!^{\prime\prime}1$.

\subsection{Dust Emission Surveys}
The first major survey of the disks around T-Tauri stars
was performed with the IRAM Plateau de Bure interferometer
at 110~GHz \cite{dutrey96},
with follow-up observations at 230~GHz \cite{guilloteau97}.
This survey produced the very important first measurements 
of well-resolved sizes for a sizable subset of the observed 
sample. Figure~\ref{fig:dutrey_pdbi} shows a set of 
high resolution 230~GHz images of sources in Taurus.
Careful modeling of the observations implied 
disks around single stars with large outer radii, $R> 150$ AU,
and rather shallow radial surface density profiles, $\Sigma\propto r^{-p}$, 
$p<1.5$.  These shallow surface density profiles imply that only a 
small fraction of the disk mass is located in the inner disk, a
difference from the conventional assumptions of Solar Nebula models.
The uniform nature of such surveys allows for investigation of correlations 
between the derived disk parameters and various evolutionary markers. 
An intriguing result from a similar survey using the Nobeyama Millimeter 
Array \cite{kitamura02} is that the disk outer radius increases with 
decreasing H$\alpha$ luminosity, perhaps due to radial expansion with age, 
as expected in accretion disks by transfer of angular momentum.

\subsection{The Angular Resolution Frontier}
The highest resolution at millimeter wavelengths comes from the Very
Large Array augmented by the link to the Pie Town VLBA antenna,
attaining a resolution of 30 milliarcseconds at 43 GHz, albeit with
limited brightness temperature sensitivity \cite{wilner02}.
Figure~\ref{fig:twhya_vla} shows VLA images of TW Hya at 43~GHz at two
different resolutions obtained with different visibility weighting
schemes {\cite{wilner00}. The lower resolution image in the left panel
has a $\sim0.\!\!^{\prime\prime}6$ (56 AU) beam that emphasizes the
spatially extended low brightness emission. The region of detectable
43~GHz emission at this resolution is $\sim100$~AU in diameter, and a
fainter halo extends to larger distances.  The image in the right
panel has a $\sim0.\!\!^{\prime\prime}1$ (5.6 AU) beam.  Little
detectable emission is visible from TW~Hya at this scale.  Only the
orders of magnitude improvement in sensitivity and resolution from the
SKA will allow imaging at substantially higher resolution.

A comparison of the thermal dust emission images of the TW Hya disk
with scattered light images is illuminating.  In the optical, the best 
angular resolution and image quality comes from the Hubble Space Telescope.
Figure~\ref{fig:twhya_hst} shows the Hubble images of TW Hya from STIS (using
coronography) and WFPC~2 (using PSF subtraction), both of which dramatically 
reveal the full extent of this nearly face-on disk in scattered light
\cite{schneider01}.
These images also emphasize the unavoidable problems of probing the 
inner regions of protoplanetary disks in scattered light, since the star 
dominates and must be blocked or removed to very high accuracy to extract 
the faint disk signal, a process that becomes exceedingly difficult 
in the immediate neighbourhood of the central star.

\begin{figure}
\begin{center}
\includegraphics[scale=0.5]{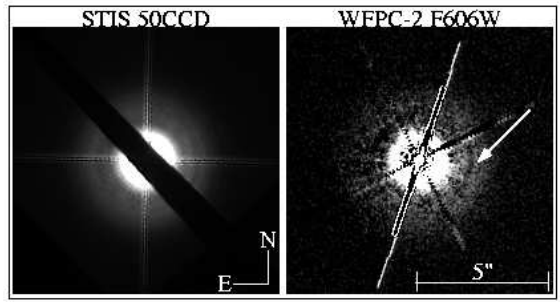}
\end{center}
\vskip -1.0cm
\caption{
The TW Hya disk imaged in scattered light from the STIS and WFPC~2 
instruments on the Hubble Space Telescope (from \cite{schneider01}).
The inner disk regions are inaccessible at these wavelengths because
of the bright central star.
}
\label{fig:twhya_hst}
\end{figure}

\subsection{Physical Models}
Physical models of disk structure and evolution are
becoming more sophisticated as more physics is 
incorporated and better computational methods are implemented.
Several groups have developed self-consistent treatments of disks
around young stars in radiative and hydrostatic equilibrium, employing
radiative transfer schemes with various degrees of complexity
(e.g. \cite{dalessio01,dullemond03}).
An important trend is to interpret high angular resolution data
in the context of these models.
Figure~\ref{fig:dalessio01} shows the results of one such calculation,
the irradiated accretion disk model of D'Alessio et al. (2001)
\cite{dalessio01}. 

\begin{figure}
\begin{center}
\includegraphics[scale=0.36]{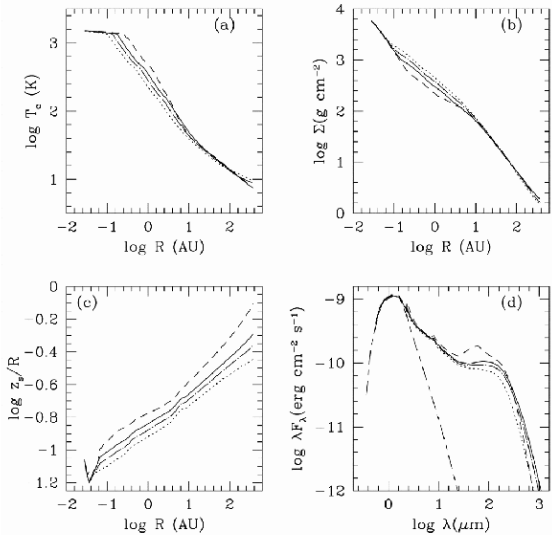}
\end{center}
\vskip -1.0cm
\caption{Results of a self-consistent accretion disk calculation for
a typical T Tauri star (from \cite{dalessio01}).  The panels show the 
radial distribution of (a) mid-plane temperature, (b) mass surface density, 
(c) height of the irradiation surface divided by radius. 
Panel (d) shows the spectral energy distributions for a pole-on viewing
geometry (the central star is indicated by the long-short dashed line).
The four curves in each panel correspond to different assumptions about 
dust grains in the model, which start to depart significantly 
within the inner disk regions that can be imaged only by the SKA. 
}
\label{fig:dalessio01}
\end{figure}

The basic features of the radial dependences that arise in this model
are readily explained. For steady accretion, the surface density is
given by $\Sigma = \dot{M}/{3\pi\nu}$ away from the boundaries, where
$\dot{M}$ is the accretion rate, and $\nu = \alpha c_s H$ is the
kinematic viscosity parameterized by a local velocity (the sound
speed, $c_s$), scale length (the scale height, $H$) and a
dimensionless parameter ($\alpha<1$).  The surface density may be
expressed as $\Sigma\propto (r^{3/2}T_m)^{-1}$, where $T_m$ is the
local mid-plane temperature whose value is determined by the balance of
heating by stellar irradiation and viscous processes, and cooling
through radiative losses. For a disk with heating dominated by stellar
irradiation, the flaring of the outer parts of the disk tends to drive
the temperature distribution to $T_m\propto r^{-1/2}$, except in the
innermost regions where the disk becomes optically thick to its own
radiation, and $\Sigma\sim r^{-1}$.

The existing resolved images of the TW Hya disk at 43~GHz
and at 345~GHz match this model very well \cite{qi04}.
The middle panel of Figure~\ref{fig:twhya_vla} shows the result of imaging 
the model brightness distribution at these two angular resolutions using 
the same visibility sampling as the VLA observations. The lower panel of 
Figure~\ref{fig:twhya_vla} shows the residual images obtained by 
subtracting the model visibilities from the VLA data and then imaging 
the difference with the standard algorithms; there are no significant
residuals.

The increasing sophistication of the physical models is leading 
to progress on many fronts. For example, a consequence of the modest 
brightness of the TW Hya disk observed within 5~AU is that the surface 
density is unlikely to be very much higher in this region of the disk 
than expected from an $r^{-1}$ extrapolation to smaller radii.
A region of low viscosity and concomitant high surface density in the 
inner disk appears to be required by some mechanisms for accretion,
planet formation, and also planet migration. For example, the layered
accretion of Gammie (1996) \cite{gammie96} piles up accreting mass in a 
``dead zone'' 
where the magneto-hydrodynamic instability does not operate very effectively.
For TW Hya, there is no evidence for any such substantial mass reservoir 
at these small radii. Indeed, the existing data suggest a mass deficit,
which may be related to a deficit of mid-infrared emission and 
a possible large central hole cleared out by a developing protoplanet
(see \S\ref{sec:gaps}).
On the large scale, if the accretion disk model for TW Hya is valid, 
then the fact that the surface density normalization based on the 
mass accretion rate (estimated from ultraviolet excess) and viscosity 
parameter $\alpha$ ($\sim0.01$) agrees with the millimeter result 
provides indirect support for the adopted mass opacity law.
	
Figure~\ref{fig:dalessio01} shows that the physical structure of the
disks start to depart from general power law behaviors in the innermost 
regions, where the models become dependent on several parameters that 
are poorly constrained by existing observations. The differences among 
the four disk models shown in each panel of 
Figure~\ref{fig:dalessio01} result from different assumptions about 
the grain size distribution. The inner regions that are affected
will be accessible to the SKA, and these assumptions may be tested.

\section{Planet Signatures: Disk Gaps}
\label{sec:gaps}
An important theoretical prediction is that planets in disks, once
they achieve sufficient mass, will interact with the disk material and
open wide radial gaps -- large regions of low surface density -- around
their orbits \cite{lin79}.  The formation of gaps could be significant
in limiting the growth of giant planets \cite{bryden99}. As noted in
\S\ref{sec:background}, the tidal interaction between a giant
protoplanet and its disk can result in orbital migration
(e.g.~\cite{lin86}), and this process affects the orbital distribution
of giant planets (and perhaps also their survival if migration is not
halted.)  Because the gaps opened by the protoplanets are much larger
than the protoplanets themselves, this gap imprinted on the disk may
be the best observational signature of protoplanet formation.

\subsection{Why Gaps?}
The basic idea behind gap formation is that an orbiting protoplanet 
tidally interacts with its parent disk. Material exterior to the planet 
gains angular momentum from the planet and moves to larger radii, while 
material interior to the planet loses angular momentum and moves to smaller
radii, creating a gap. Viscous processes tend to act against tidal forces 
and fill in the developing gap. Whether or not the tidal forces 
overcome the viscous processes depends on the mass ratio $q$ of the planet 
and the star, and the Reynolds number, ${\cal{R}} = \Omega r^2/\nu$,
where $\Omega$ is the angular frequency, $r$ is the radius, and 
$\nu$ is the kinematic viscosity (often parameterized using the 
$\alpha$ prescription). If $q>{\cal{R}}^{-1}$, then a gap forms. 

Figure~\ref{fig:gaps} shows the result of a representative calculation 
where the disk surface density is perturbed and develops a gap
\cite{bryden99}.  In this hydrodynamic simulation, the mass ratio,
$q$, is $10^{-3}$, and trailing spiral density waves are excited near 
the inner and outer Lindblad resonances of the protoplanet. 
Shock dissipation of waves deposit angular momentum, and the disk material
moves away from the protoplanet orbit. The gap width is $\sim20$\% of
the orbital radius, or $\sim1$~AU for a Jupiter at 5~AU.
More sophisticated simulations are now being undertaken that include 
prescriptions for magnetohydrodynamical turbulence 
\cite{papaloizou04}
and fully three dimension hydrodynamics \cite{bate03}, and the results 
are qualitatively very similar. For typical initial conditions, planets 
with masses as low as $\sim0.1$ Jupiter mass open significant gaps, though 
lower mass planets launch spiral waves and produce potentially detectable 
large scale disturbances.  Since the physical conditions of the inner disk 
are poorly constrained, and the sources of disk viscosity are even more
poorly understood, such results will undoubtedly undergo revision.
In principle, the properties of the protoplanets ultimately may be 
constrained in detail by measurements of the gap locations and widths, 
and the amount of residual material within the gaps.

\begin{figure*}
\begin{center}
\includegraphics[scale=0.75]{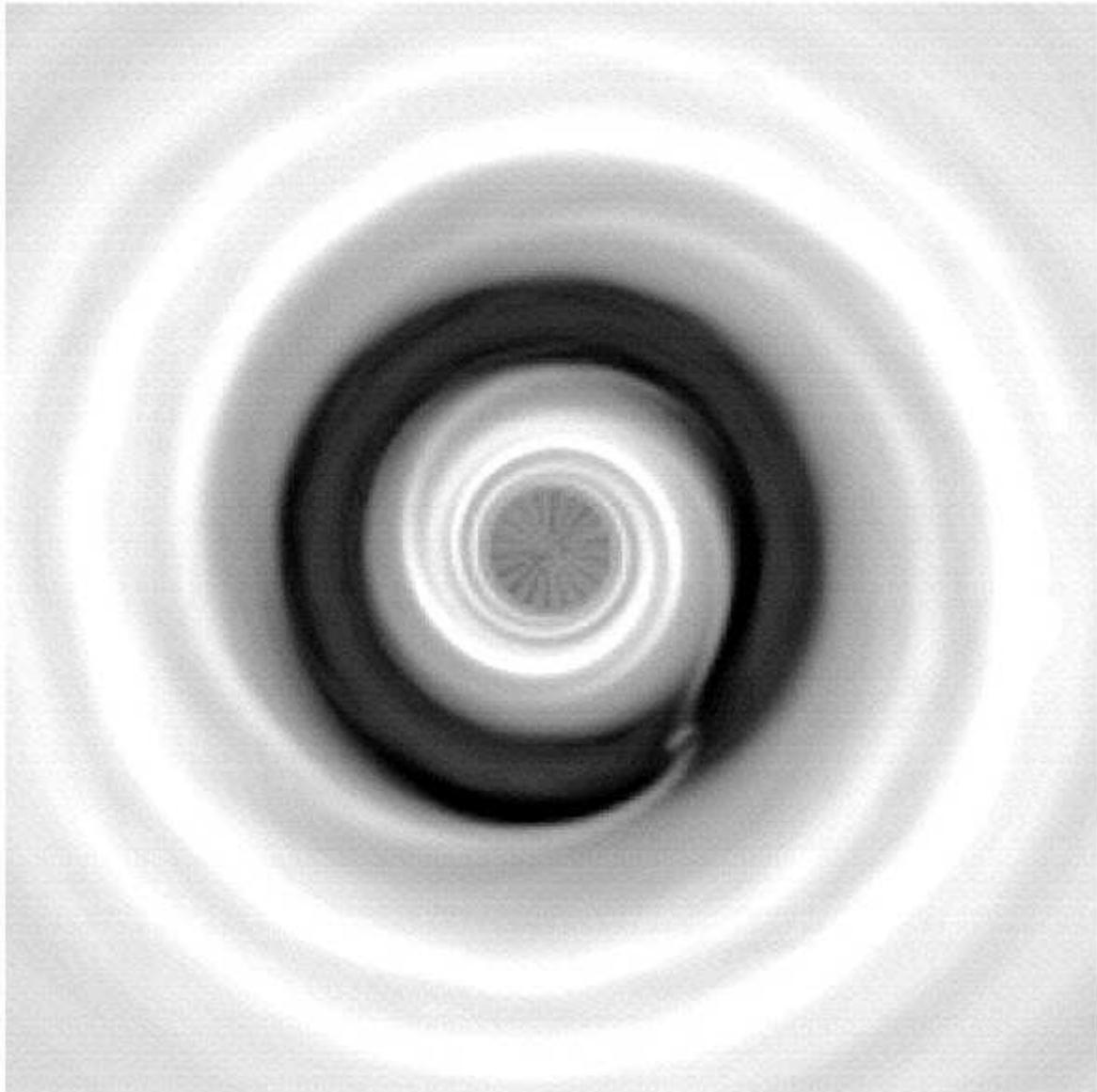}
\end{center}
\vskip -0.75cm
\caption{A numerical simulation of a Jupiter mass planet within a disk 
that opens up a gap in the disk through tidal interaction 
(from \cite{bryden99}). 
The dark region in this surface density plot is nearly devoid of 
disk material.  The planet is located within the dark region, in a small
accretion stream. While the planet itself is extremely difficult to
detect at any wavelength, the wide gap provides a strong marker of 
its presence, and the gap structure might be used to constrain 
its properties. 
}
\label{fig:gaps}
\end{figure*}

\subsection{Evolutionary Issues}
The subsequent evolution of a disk with a gap induced by a protoplanet 
may include accretion of material inside the protoplanet orbit to produce 
an inner evacuated hole, if gas from the outer disk is prevented from 
accreting across the gap. The formation of Jupiter in our Solar System might 
have 
prevented the outer disk gas from reaching interior regions and allowed 
the inner disk gas to fully accrete, leaving behind solid planetesimals 
to form terrestrial planets. But if the planet were to lose angular 
momentum due to torques from a sufficiently massive outer disk, then the
Jupiter would migrate inward and sweep away any interior rocky planets.
Observations are needed to inform the frequency of such disparate outcomes 
of the planet building process. 

\subsection{Observing Gaps}
Disk gaps may manifest themselves in spectral energy distributions as
the deficit of disk material over a limited range of radii results in
less emission in a limited wavelength range. A gap close to the star
might result in a deficit of mid-infrared emission, which is otherwise
dominated by close-in material. Some T-Tauri stars, for example TW
Hya, GM Aur, and CoKu Tau 4, have mid-infrared deficits that indicate
the inner disks have been cleared out to radii of a few AU or more.
Less dramatic perturbations in the disk structure, such as spiral
waves generated by a low mass planet, will have more subtle effects on
the spectral energy distribution, and it is likely they will be
revealed only by direct imaging of surface density structure.

If the predicted gaps are present in the inner disk, SKA images of
dust emission will show them. Such gaps, if they are not devoid of
gas, may have spectral signatures at infrared wavelengths, for example
emission in the CO fundamental lines at 4.7~$\mu$m, which should be
detectable with large optical telescopes \cite{najita03}.  Such gaps
may also slightly modify the visibilities obtained by interferometers
operating at infrared wavelengths, though teasing out the influence of
the gaps from other model parameters may be impossible \cite{wolf02}.
The SKA will be unique in {\em directly imaging} the gaps, which will
avoid any ambiguities in the interpretation of unresolved
observations, or highly model dependent results.

It is worth noting that ALMA will not have sufficient angular
resolution to probe structure in the habitable zone of the nearby
disks. In its most extended antenna configuration, the angular
resolution will be only 20 milliarcseconds at 345~GHz (3~AU at 150
pc). In an optimistic scenario, ALMA will be able to reveal the gap
created by a Jupiter orbiting at 5 AU \cite{wolf02}. The high dust
opacity at these high frequencies also will be problematic for
investigating the inner disk.  Observations of inner disk structure
must be made in the optically thin regime at lower frequencies using
the SKA.

\section{Proper Motions}
Observations of dust emission do not provide any kinematic information
directly, but the orbital timescales are sufficiently short that 
{\em synoptic} studies can follow the proper motions of mass concentrations 
and structures within the disk. At a characteristic radius of 
1~AU from a T-Tauri star of $\sim1$~M$_{\odot}$, the full orbital period 
is $\sim1$ year. Thus a series of images of several disk systems with 
identifiable features could be made at intervals of approximately a month, 
thereby providing good coverage of one orbit or more. These short timescales 
will permit the tracking of secular changes in disk structure related to 
the presence of planetary mass bodies. A wide spectrum of resonant 
interactions are possible that may produce dynamical signatures such as
spiral waves. In addition, proper motions of features monitored over 
time will allow for clear discrimination of faint disk structures from any 
sources of confusion that might be present in the extragalactic background.

\section{Summary}
The SKA will play a pivotal role in understanding terrestrial planet
formation by imaging thermal dust emission from the inner regions of
protoplanetary disks around nearby young stars. The observations will
probe the dense disk material at size scales of 1~AU, commensurate
with the orbit of the Earth around the Sun. SKA imaging will reveal
structures produced by embedded protoplanets whose motions can be
tracked over time in a rich and representative sample of currently
forming Solar Systems.  Future optical and infrared planet searches
may reveal terrestrial planets that could harbor life, but it will be
left to the SKA to tell us why they are there and why we are here.

\bigskip
D.J.W. acknowledges support from NASA Origins of Solar Systems Program 
grant NAG5-11777. 
Thanks to 
Sean Dougherty, 
Lewis Knee and others in the 
SKA Working Group on the Life Cycle of Stars
for many valuable contributions.

\end{document}